# In-flight performance and calibration of the Infrared Array Camera (IRAC) for the Spitzer Space Telescope

J. L. Hora[*,a], G. G. Fazio[a], L. E. Allen[a], M. L. N. Ashby[a], P. Barmby[a], L. K. Deutsch[a], J.-S. Huang[a], M. Marengo[a], S. T. Megeath[a], G. J. Melnick[a], M. A. Pahre[a], B. M. Patten[a], H. A. Smith[a], Z. Wang[a], S. P. Willner[a], W. F. Hoffmann[b], J. L. Pipher[c], W. J. Forrest[c], C. W. McMurty[c], C. R. McCreight[d], M. E. McKelvey[d], R. E. McMurray[d] Jr., S. H. Moseley[e], R. G. Arendt[e], J. E. Mentzell[e], C. T. Marx[e], D. J. Fixsen[e], E.V. Tollestrup[f], P. R. Eisenhardt[g], D. Stern[g], V. Gorjian[g], B. Bhattacharya[h], S. J. Carey[h], W. J. Glaccum[h], M. Lacy[h], P. J. Lowrance[h], S. Laine[h], B. O. Nelson[h], W. T. Reach[h], J. R. Stauffer[h], J. A. Surace[h], G. Wilson[h], E. L. Wright[i]

[a]Harvard-Smithsonian Center for Astrophysics, 60 Garden St., MS-65, Cambridge, MA 02138; [b]Steward Observatory, Univ. of Arizona, 933 North Cherry Ave., Tucson, AZ 85721; [c]Dept. of Physics and Astronomy, Univ. of Rochester, Rochester, NY 14627; [d]NASA Ames Research Center, Moffett Field, CA 94035; [e]NASA Goddard Space Flight Center, Greenbelt, MD 20771; [f]IfA, Univ. of Hawaii at Manoa, 2680 Woodlawn Dr., Honolulu, HI 96822; [g]JPL, MS 264-767, 4800 Oak Grove Dr., Pasadena, CA 91109; [h]Spitzer Science Center, California Institute of Technology, 1200 E. California Blvd., Pasadena, CA 91125; [i]Dept. of Physics and Astronomy, Univ. of California at Los Angeles, P.O. Box 951562, Los Angeles, CA 90095;

## ABSTRACT

The Infrared Array Camera (IRAC) is one of three focal plane instruments on board the *Spitzer Space Telescope*. IRAC is a four-channel camera that obtains simultaneous broad-band images at 3.6, 4.5, 5.8, and 8.0 µm in two nearly adjacent fields of view. We summarize here the in-flight scientific, technical, and operational performance of IRAC.

## 1. INTRODUCTION

The IRAC instrument[1,2] is a four-channel camera on board the *Spitzer Space Telescope*[3,4] that can obtain simultaneous 5.2×5.2 arcmin images in broad-band filters centered at wavelengths of 3.6, 4.5, 5.8, and 8 µm. Two nearly adjacent fields of view are imaged in pairs (3.6 and 5.8 µm; 4.5 and 8.0 µm) using dichroic beamsplitters. All four detector arrays[5] in the camera are 256×256 pixels in size with a pixel size of approximately 1.22 arcsec. The two short wavelength channels use InSb detector arrays[6] and the two longer wavelength channels use Si:As detectors[7,8]. The camera has an internal calibration subsystem that can illuminate the detectors through the IRAC optics to measure the system responsivity, and a separate flood calibrator system that directly illuminates the detectors. Only the flood calibrators are used in flight.

The first 90 days of the mission were reserved for the In-Orbit Checkout (IOC) and Science Verification[9] (SV) of the instruments, telescope, and spacecraft. During this period the telescope cooled from ~300K at launch to ~5.5K, and the telescope was focused[10,11]. The IRAC observations to confirm functionality, focus, and calibrate the instrument during this period were planned before launch and for the most part executed as scheduled except for rearrangement of some tasks and the addition of some new tests to investigate anomalies. Several early release observations were obtained during the IOC/SV period, and the first IRAC science campaign began on 2003 Dec 1.

---

[*] jhora@cfa.harvard.edu, http://cfa-www.harvard.edu/irac

During nominal operations, observations are planned using the Astronomical Observing Template (AOT)[12] in the *Spitzer* Planning Observations Tool (SPOT). When planning is complete, the AOTs are transferred to the *Spitzer* Science Center (SSC) with SPOT, where they are processed into Astronomical Observation Requests (AORs) which are then scheduled and executed. The instruments are operated in "campaigns" (typically 7-10 days long for IRAC) during which one instrument is on and performing a pre-planned sequence of spacecraft commands and observations. As of the end of May 2004, IRAC has had 8 science campaigns and a total of over 2500 hours of operation since launch. The first science results from *Spitzer* will be published later this year in a special issue of the Astrophysical Journal Supplement Series. Results were also presented at the January and May 2004 American Astronomical Society meetings.

## 2. SENSITIVITY

The point source sensitivities for the frame times available in the IRAC AOT are shown in Table 1. For each frame time and channel, the 1σ sensitivity in µJy is given for the low background case (near the ecliptic poles). The values were calculated based on the sensitivity model developed before launch[13], but using measured in-flight values for the read noise, pixel size, noise-pixels, background, and total system throughput. The noise-pixel value is the equivalent number of pixels that contribute to noise in the analysis when an image is spatially filtered for optimum faint point-source detection[14]. The numbers were compared to observations at several frame times to confirm the validity of the calculations. The values in Table 1 are close to pre-flight predictions except for two factors: the lower (better) than expected noise-pixel values for all channels, and the lower (worse) throughput in channels 3 and 4. These effects almost cancelled each other out in channels 3 and 4, and allowed IRAC to meet its required sensitivity of 0.92, 1.22, 6, and 9 µJy (1σ, 200 sec) for channels 1-4, respectively.

Table 1. IRAC Point Source Sensitivity (1σ µJy, low background)

| Frame Time (sec) | 3.6 µm | 4.5 µm | 5.8 µm | 8.0 µm |
|---|---|---|---|---|
| 200 | 0.40 | 0.84 | 5.5 | 6.9[†] |
| 100 | 0.60 | 1.2 | 8.0 | 9.8[†] |
| 30 | 1.4 | 2.4 | 16 | 18 |
| 12 | 3.3 | 4.8 | 27 | 29 |
| 2 | 32 | 38 | 150 | 92 |
| 0.6[*] | 180 | 210 | 630 | 250 |
| 0.4[‡] | 86 | 75 | 270 | 140 |
| 0.1[‡] | 510 | 470 | 910 | 420 |
| 0.02[‡] | 7700 | 7200 | 11000 | 4900 |

[*]high dynamic-range mode
[†]using multiple 50 sec frames
[‡]subarray mode

The noise measurements in channels 3 and 4 scale as expected for long series of integrations with the inverse square root of time, to the limit measured (15000 seconds)[2]. Channels 1 and 2 deviate from this rule in approximately 2000 seconds, which indicates that these channels are approaching the source completeness confusion limit. The deviation occurs at about 1 µJy (5σ), which is slightly fainter than the pre-flight predictions for the IRAC confusion limit[15].

## 3. IRAC IMAGE QUALITY

### 3.1. PSF and focus

The IRAC camera itself provides diffraction-limited imaging internally with wavefront errors of $< \lambda/20$ in each channel. The image quality in flight is limited primarily by the *Spitzer* telescope, which is estimated to be diffraction-limited (rms wavefront error $< \lambda/14$) at 5.5µm[11]. The PSF was monitored throughout the early mission while the telescope cooled down, and while the secondary mirror was refocused in two moves to achieve a near-optimal focus[10,16]. The telescope temperature varies between approximately 5.5 to 12 K during a typical IRAC campaign with no detectable change in the PSF. Table 2 shows some properties of the IRAC point spread function (PSF). These numbers were derived from in-flight measurements of bright stars after the focus operation was complete.

Table 2. IRAC Image Quality

| Channel | Noise-Pixels (mean) | FWHM (mean; arcsec) | FWHM of centered PSF (arcsec) | Central pixel flux (peak; %) | Pixel Field of View (arcsec) | Max. distortion (pixels relative to square grid) |
|---|---|---|---|---|---|---|
| 1 | 7.0 | 1.66 | 1.44 | 42 | 1.221 | 1.3 |
| 2 | 7.2 | 1.72 | 1.43 | 43 | 1.213 | 1.6 |
| 3 | 10.8 | 1.88 | 1.49 | 29 | 1.222 | 1.4 |
| 4 | 13.4 | 1.98 | 1.71 | 22 | 1.222 | 2.2 |

There are two columns for the PSF full width at half-maximum (FWHM). The mean FWHM is from observations of a star at 25 different locations on the array. The FWHM for the "centered PSF" is for cases where the star was most closely centered in a pixel. The fifth column in Table 2 is the percentage of the flux in the central pixel for a source that is well-centered on a pixel. The flux in the central pixel for a random observation will be lower, because the PSF of the telescope is rather undersampled at the IRAC pixel scale, except in channel 4. The reconstructed center-of-field PSF is shown in Figure 1.

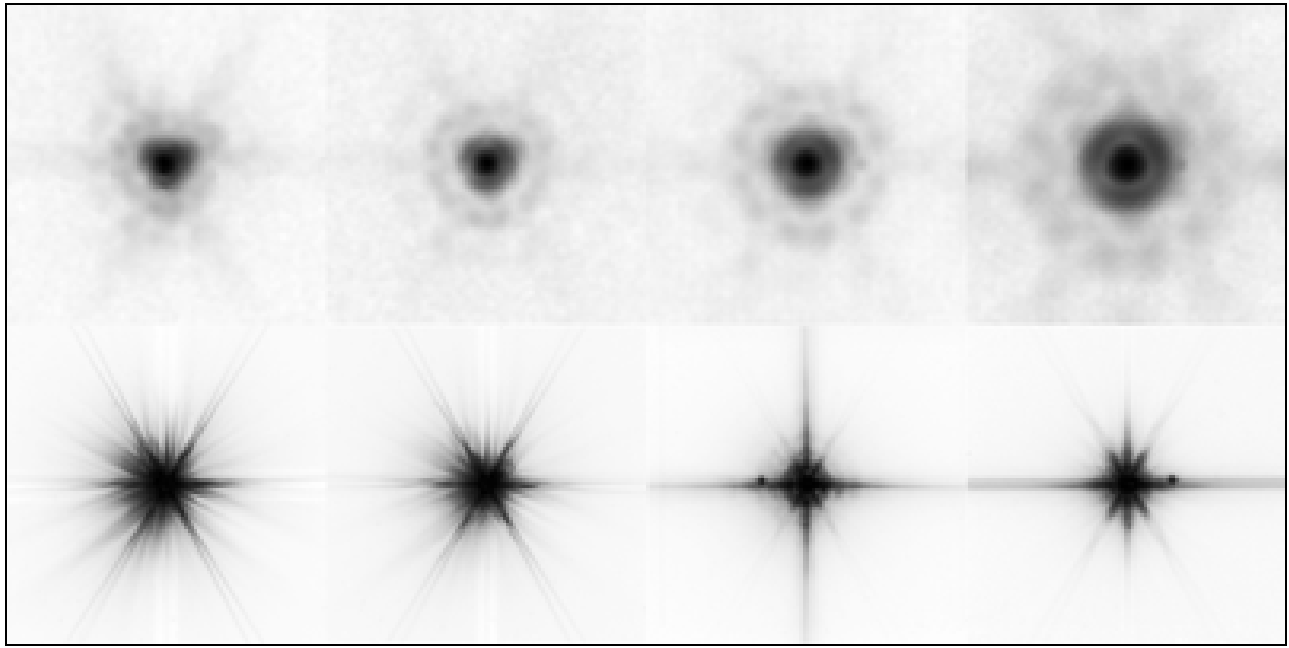

**Figure 1.** The IRAC PSF, derived from observations of bright stars. The two rows have channels 1 – 4 from left to right. A logarithmic scaling is used, with highest intensity black and lowest white. The images in the top row were constructed from many separate exposures, using the drizzle algorithm on a pixel scale ¼ the linear size of the IRAC instrument pixels. The top row shows the inner 30 arcsec of the PSF, scaled to show the core and Airy rings. The bottom row shows the extended PSF in a 5 arcmin square box. The six diffraction spikes are visible, extending horizontally and in an X pattern centered on the core, and each spike splits in two outside of the central region. Some other image and array artifacts are also visible. In channel 2, a ghost image is visible as a small arc to the upper right of the core. In channels 3 and 4, the ghosts are seen displaced about 45 arcsec to the left (channel 3) or right (channel 4) of the star (the small spots). Banding is visible in channels 3 and 4 as the enhancement of the horizontal or vertical bands that extend from the center to the edges of the array.

## 3.2. IRAC pipeline-processed data quality

The IRAC data are processed by the SSC to produce the Basic Calibrated Data (BCD), which are single IRAC frames that have most of the instrumental signatures removed and have been calibrated to units of MJy/sr. The main steps are to remove the detector dark frame pattern, linearize and flatten the data, and apply the flux calibration. See the *Spitzer* Observer's Manual[*] for a full description of the pipeline processing steps. Some example BCD frames are shown in Figure 2. These are individual 12 sec frames from the region near NGC 7129 taken as Early Release Observations, and are publicly available from the SSC. Figure 3 shows a mosaic constructed from the entire dataset illustrating what is possible after cosmic ray and bad pixel rejection by combining dithered images.

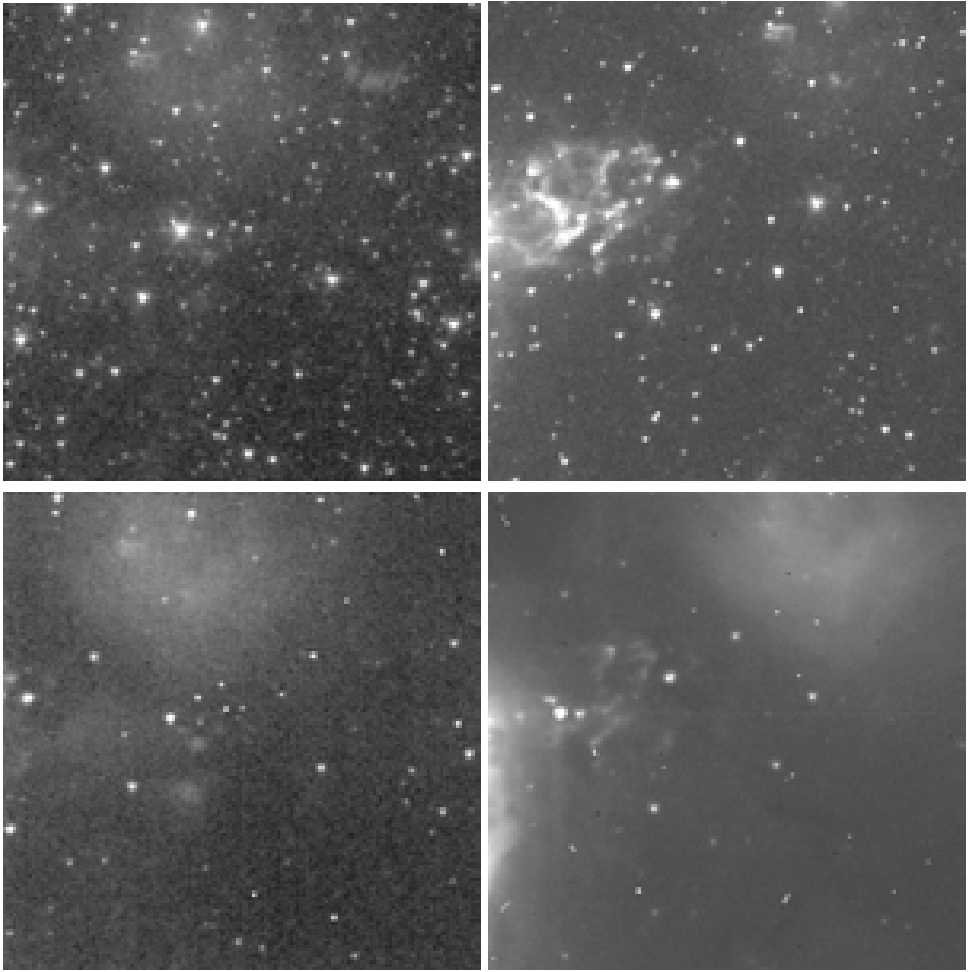

**Figure 2.** Basic Calibrated Data (BCD) images from the IRAC pipeline for a region in NGC 7129, all 12 sec frames. Channels 1-2 are in the top row, left and right, and channels 3 – 4 are in the bottom row, left and right. Each image is a 5.2×5.2 arcmin region on the sky. The channel 1 and 3 images are simultaneous images of the same field, and the channel 2 and 4 images are of a field that is slightly offset from that shown in channels 1 and 3. The region viewed is near the tip of the green feature just to the right of center in Figure 3. There are a few bright pixels due to cosmic rays and bad pixels, but most of the white spots are real sources. The number of visible sources decreases at longer wavelengths, since most of the objects are normal stars following roughly the Rayleigh-Jeans portion of the blackbody spectrum. There is some faint diffuse nebulosity in the upper part of the frames which appears red in Figure 3, and some highly structured nebulosity on the left side of the images which appears brightest in channel 2, and therefore green in Figure 3. The bright emission in channel 2 is likely due to shocked-excited molecular emission caused by outflows from the young stars.

---

[*] http://ssc.spitzer.caltech.edu/documents/som/

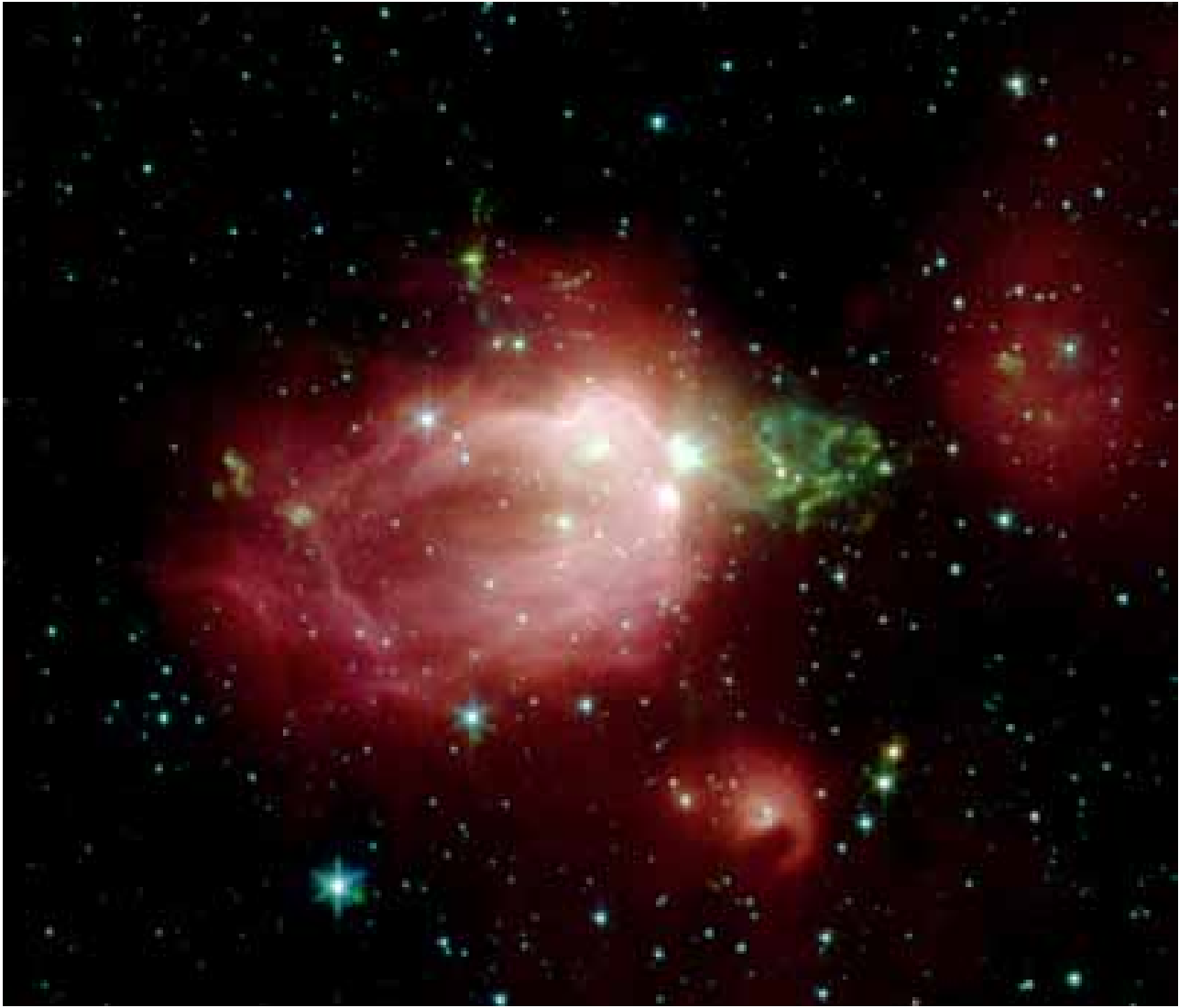

**Figure 3.** The reflection nebula NGC 7129 and associated stellar cluster imaged by IRAC[17]. In this four-color composite, emission at 3.6 μm is depicted in blue, 4.5 μm in green, 5.8 μm in orange, and 8.0 μm in red. The image size is approximately 14×12 arcmin. The effective total integration time at each position of the mosaic is approximately 40 seconds in each channel (4 dithers per mosaic position).

### 3.3. Distortion

There is a small amount of distortion over the IRAC FOV. The maximum distortion in each IRAC band is < 2.2 pixels (compared to a perfectly regular grid) over the full FOV. The absolute IRAC positional accuracy of the BCD, refined using 2MASS source matches, is 0.2 arcsec. The distortion results in a slightly different size of the sky being viewed by each pixel (see Figure 4), and in conjunction with the method of measuring the in-flight flat fields, requires a correction to the point source photometry that is a function of the source position in the frame. The flat field is measured by observing two regions with different zodiacal backgrounds and devoid of bright point sources, one near the ecliptic pole and one near the ecliptic plane. The normalized difference between the two images is the flat field which will accurately correct the extended emission when divided into a science frame. However, the flat-fielding method assumes that for a heavily dithered sky field, the counts/pixel should be constant over the entire medianed image. Since the area of each pixel varies systematically over the FOV, that assumption is invalid and therefore the resultant flat field is imperfect for point source photometry. The photometry for point sources must be corrected by the area factor of the pixels within the measurement aperture. The largest corrections are ± 1.24%, 1.6%, 1.2%, and 1.9% for channels 1 – 4, respectively.

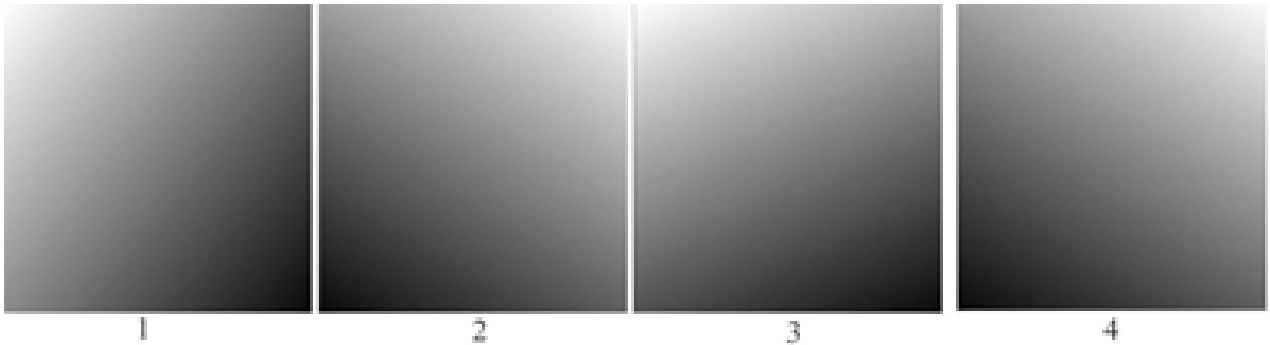

Figure 4. The differences in pixel area on the sky over the array, due to the optical distortion in each of the bands. The largest positive difference from the average area is white, while the largest negative difference from the average is black in these images.

### 3.4. Scattered and Stray Light

IRAC images exhibit some stray and scattered light artifacts from sources nearby the IRAC fields of view. The sections below show some examples of these features. More discussion is presented on an SSC web page[†].

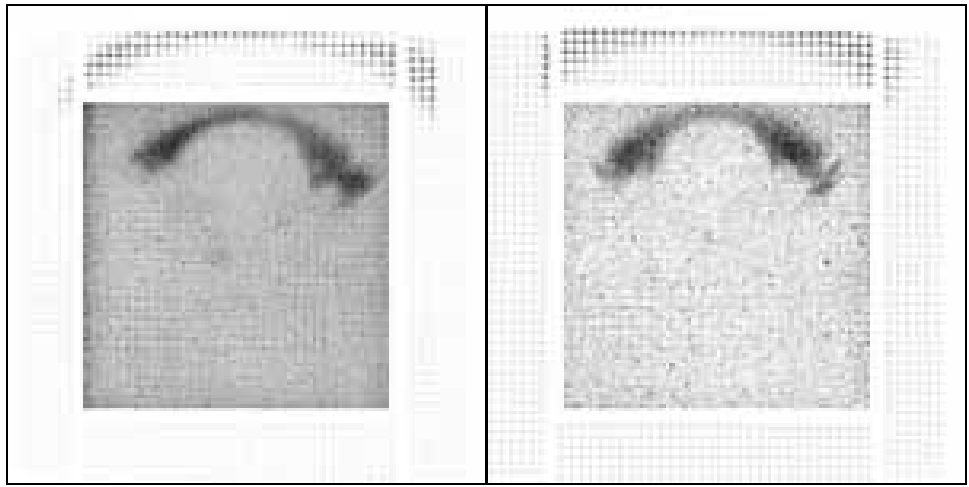

**Figure 5.** Average of frames taken during a stray light test. Channel 1 is on the left and channel 2 on the right. Bright point sources were placed at each of the locations outside the field of view indicated by "+" symbols, and the symbols are shaded to represent the total intensity of the stray light seen from each location (black is higher intensity in these images). The "butterfly" is the result of the sum of stray light from all sources in the vulnerable region.

### 3.4.1. Diffuse stray light

Diffuse stray light from outside the IRAC FOV is scattered into the active region of the IRAC detectors in all four channels. The problem is significantly worse in channels 1 and 2 than in channels 3 and 4. Stray light has two implications for observers. First, diffuse background light, when scattered into the arrays, is manifest as additions to the flat fields when they are derived from sky observations. In channels 1 and 2 the stray light pattern due to diffuse sources resembles a "butterfly", shown in Figure 5.

---

[†] http://ssc.spitzer.caltech.edu/irac/scatt.html

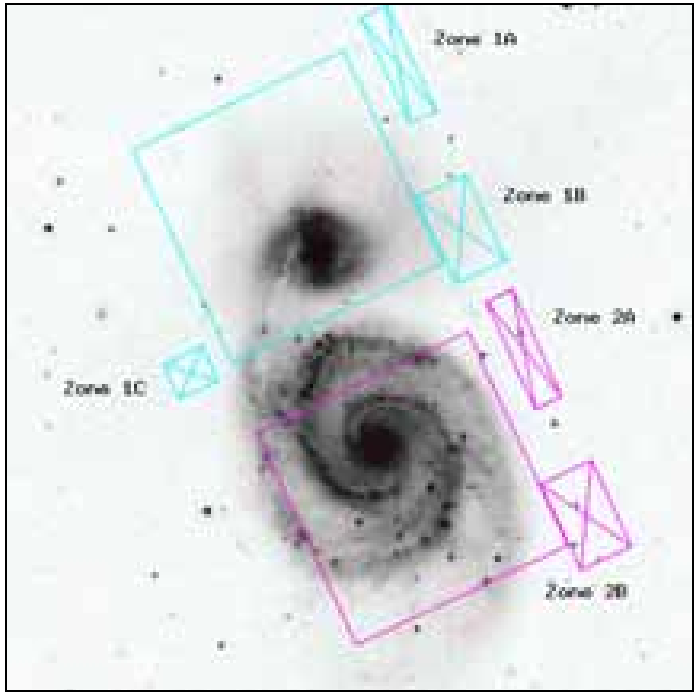

**Figure 6.** Locations of IRAC stray light avoidance zones as seen in the SPOT visualizations, superposed on an optical image of the M51 galaxy. The top large square represents the field of view of channels 1/3 and the bottom square channels 2/4. The stray light avoidance zones are indicated by the rectangles with

Second, patches of stray light from discrete sources can show up as spurious sources in the images. Discrete sources falling on certain areas of the focal plane assembly (FPA) can scatter off the holes in the array covers and onto an array. In addition, some of the light that hits the edges of the relatively thick detector dice in channels 3 and 4 is redirected into the arrays and appears as multiple smeared images. Diffuse light that hits the edges of the channel 3 and 4 detector dice forms a "tic-tac-toe" pattern in the image. The diffuse stray light scales with the zodiacal background level, which is the source for the flat fields, so the stray light pattern contaminates the raw data used for the flats. The SSC pipeline attempts to remove diffuse scattered light by scaling a template of the stray light pattern by the estimated zodiacal background level at that position and subtracting it from the data, flat-fields and darks. This has been effective in removing most of the adverse effects on the photometry that was introduced by having the scattered light in the flat fields. However, a small residual may exist in the BCD if the zodiacal background estimate for the field is not perfect.

### 3.4.2. Scattered light from stars

Stars which fall into regions that scatter light into the detectors produce distinctive patterns of scattered light on the array. Scattered light avoidance zones, which are just outside the arrays (see Figure 6), have been identified in each channel where observers should avoid placing bright stars if their observations are sensitive to scattered light. Typically, in channels 1 and 2 about 2% of the light from a star is scattered into a "splatter pattern", which has a peak intensity of about 0.2% of the light from the star. These scattered light patterns are shown in Figure 7 and Figure 8. The avoidance zones for channels 3 and 4 are a narrow strip about 3 pixels wide, 16 pixels outside of the array and surrounding it. Well-dithered observations can move the stars causing stray light patches outside of the stray light avoidance zones, which will allow the stray light patches to be rejected as outliers at the mosaicing stage.

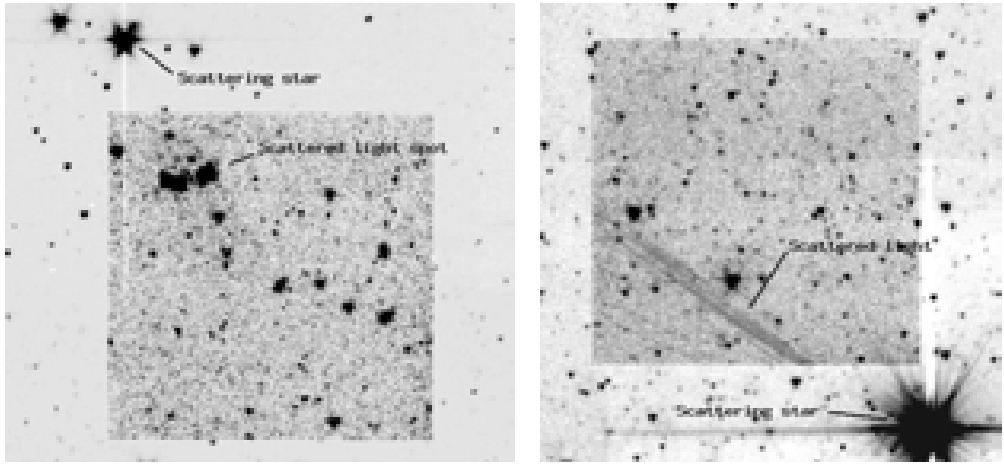

**Figure 7.** Examples of stars scattering light into IRAC channels 1 and 2. The original IRAC frames are the darker squares, they have been superposed on larger mosaics to show the relative location of the scattering star. Left – A star in zone 1C scattering light into channel 1. Right – A star in zone 2A scattering light into channel 2.

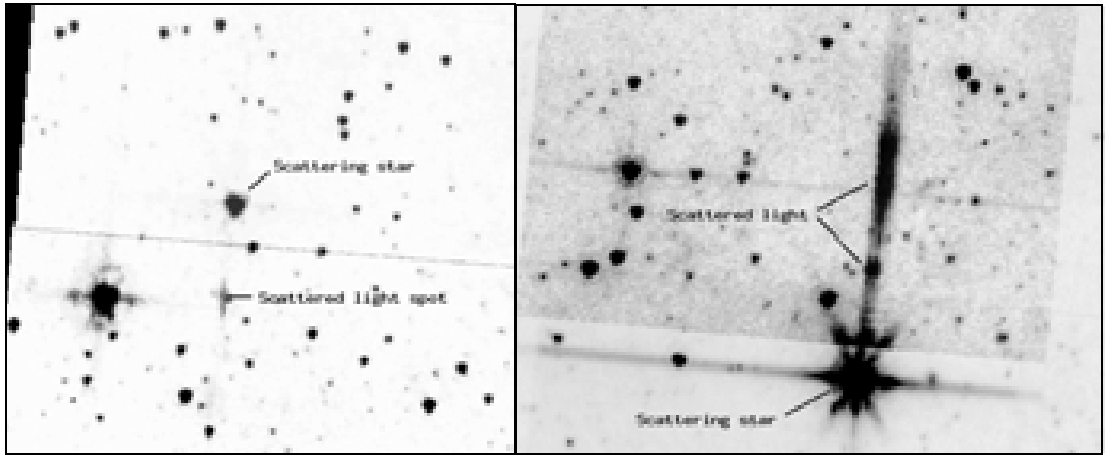

**Figure 8.** Left – Example of a star in the strip surrounding channel 3 scattering light. Right – A star in the strip surrounding channel 4 producing two scattered light spots. The peak of the first spot in from the edge of the array corresponds to a position on the "lattice" structure seen in the flat field. The second, more extended spot is only seen for the brightest scattering sources in channels 3 and 4 and probably corresponds to a second reflection of the light causing the first spot.

### 3.5. Ghost images

Ghost images are visible near very bright sources in channels 1 and 2. These ghost images are caused by internal reflections within the tilted filters or IRAC lenses. Ghost images from the filters are not in focus and typically triangular in shape. In channels 3 and 4 filter ghost images also appear for very bright (saturated) sources and are cross-shaped (+). The peak intensities of these ghosts are < 0.5% of the (unsaturated) peak intensity of the source, but they have not yet been well-characterized because ghost-producing sources are typically saturated.

A second type of ghost is an image of the telescope pupil, shown in Figure 9. These ghosts appear in specific regions near bright stars. Only a portion of the pupil image is illuminated, depending on where the star is located in the field of view. Pupil ghosts have a surface brightness roughly comparable to the diffraction spikes. Their cause is not completely understood, but could be due to internal reflections in the IRAC optics.

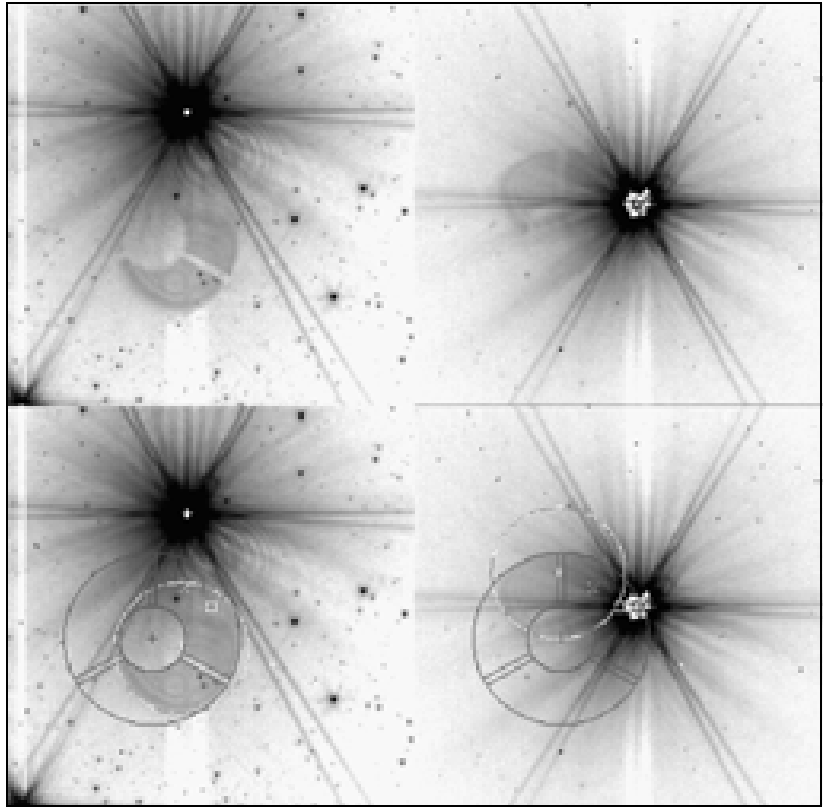

**Figure 9.** The Channel 2 pupil ghost as seen for V416 Lac (AOR 4958976; left) and Fomalhaut (right). The top row shows the images on a log scale. The bottom row shows the superimposed outline (large and small concentric circles with vanes) of the entrance pupil (i.e. the *Spitzer* primary mirror, with secondary and metering tower obscurations), and a rough estimate (smaller gray-white circle) of the illuminated portion of the pupil image. The cross ("×") indicates the pivot point for the location of the illuminated region. The square marks the center of the array.

## 4. SYSTEM THROUGHPUT

The IRAC system throughput and optical performance is governed by a combination of the system components, including the lenses, beamsplitters, filters, mirrors, and detectors. The total system throughput is assumed to be the same as the pre-flight predictions[13], based on the measurements of the optical components[18] and detector properties[6,8]. The IRAC/*Spitzer* total throughput properties are summarized in Table 3. The bandwidth is the full width of the band at 50% of the average in-band transmission. The center $\lambda$ is the wavelength that is the midpoint between the wavelengths that define the bandwidth. The average, minimum, and maximum throughput is determined in the central 90% of the bandwidth, and includes the detector QE, *Spitzer* telescope transmission (including the central obscuration), and IRAC optics. See Fazio et al.[2] for transmission plots and the isophotal wavelengths.

**Table 3.** IRAC Channel Throughput

| Channel | Center $\lambda$ (μm) | Bandwidth (μm) | Average Throughput | Minimum In-band Throughput | Maximum Throughput |
|---|---|---|---|---|---|
| 1 | 3.56 | 0.75 (21%) | 0.426 | 0.339 | 0.465 |
| 2 | 4.52 | 1.01 (22%) | 0.462 | 0.330 | 0.535 |
| 3 | 5.73 | 1.42 (25%) | 0.150 | 0.119 | 0.170 |
| 4 | 7.91 | 2.93 (37%) | 0.280 | 0.199 | 0.318 |

# 5. CALIBRATION

The overall requirement for the IRAC mission is that the system photometric responsivity be calibrated to a relative accuracy of 2% and that the absolute calibration of the data set be determined to an accuracy of better than 10%. In order to meet these requirements, it is necessary to observe calibration standards and correct for the instrumental signatures in the calibration and science data, including detector non-linearities, pixel-to-pixel sensitivity variations, and dark & bias offsets.

Many of the necessary calibrations, such as first-frame effect, linearity, and PSF characterization, were performed before launch[13] or during IOC. These instrument characteristics are assumed to be slowly varying and do not require repeated measurement on a per-campaign basis. Other measurements are repeated each campaign since they may change on shorter timescales or require many measurements over several campaigns to achieve adequate signal to noise. The routine calibration program performed each campaign includes observations of astronomical standard stars, dark frames, and flat field observations. A number of astronomical standard stars must be observed to obtain a valid absolute flux calibration[19,20]. Stars with a range of spectral indices and fluxes are observed at a number of positions across the array and many times throughout the mission to monitor any changes that may occur.

## 5.1. Calibration stars

The IRAC data are calibrated using 11 primary calibrators (PC) and (eventually) about 30 secondary calibrators (SC). The primary calibrators are located in the continuous viewing zone (CVZ) and are observed twice per campaign. The secondary calibrators are located in the ecliptic plane and are observed once during each period of autonomous observation (PAO) or about once per twelve hours. The PC's (SC's) provide measures of the long-term (short-term) baseline stability. Since the SC's are observed close to downlinks, any candidate must fall in a window diametrically opposed to the earth in the ecliptic plane. This window drifts by a full 360 degrees per year and any one candidate is visible only for a couple of campaigns at most.

The primary calibrators were chosen from ground-based observations of candidate stars[19]. The calibration stars are A dwarfs or K giant stars because these are most easily modeled. The zero magnitude fluxes in the bands were determined to be 277.5, 179.5, 116.6, and 63.1 Jy for channels 1 – 4, respectively. The primary standard stars are listed in Table 4. The stars with names starting with "NPM1" are from the Lick Northern Proper Motion Survey[21], and the stars with names starting with "KF" are from an optical/near-infrared survey of the north ecliptic pole[22].

Table 4. Flux densities (mJy) of IRAC standard stars

| Standard Star | 3.6 μm | 4.5 μm | 5.8 μm | 8.0 μm |
|---|---|---|---|---|
| NPM1p67.0536 | 844 | 483 | 325 | 185 |
| HD 165459 | 648 | 421 | 275 | 148 |
| NPM1p68.0422 | 585 | 336 | 227 | 129 |
| NPM1p64.0581 | 38.2 | 24.8 | 16.1 | 8.71 |
| NPM1p66.0578 | 141 | 82.4 | 55.5 | 31.6 |
| KF09T1 | 170 | 105 | 68.3 | 38.8 |
| NPM1p60.0581 | 38.2 | 24.8 | 16.1 | 8.70 |
| KF06T1 | 13.9 | 7.92 | 5.43 | 3.09 |
| KF08T3 | 11.8 | 7.25 | 4.73 | 2.69 |
| KF06T2 | 10.5 | 6.00 | 4.12 | 2.34 |
| 2MASS 1808347 | 6.80 | 4.43 | 2.89 | 1.56 |

The calibration appears to remain constant with time. No discernible trends are apparent within the current uncertainties. The scatter between standards is consistent with the 2%-3% uncertainty expected in the stellar models[20]. The relative calibration stability for a particular standard star over the six campaigns is in the 1%-2% range for all channels. Figure 10 shows the calibration factors derived from each star measurement plotted for science campaigns 1

– 6, which spans a period of approximately 150 days. Currently, a different flat field and dark are used for each campaign, which could be contributing to some of the scatter of the measurements. As knowledge of the darks and flats improve, the scatter will likely become smaller. The standards will continue to be observed throughout the mission to refine the calibration and monitor any long-term changes in detector responsivity that may occur.

## 5.2. Flat Field

The flat field for each channel is defined as the factors by which one must correct each pixel to give the same value at a particular flux for uniform illumination. The factors include any field-dependent optical transmission effects as well as the quantum efficiency of the pixel and output electronics gain. Flat fields are obtained from observations of the sky during normal operations. If the data are taken with an appropriate dither pattern, it is possible to relate the total response of each pixel to that of all others. Observations through the entire telescope and instrument are necessary to measure the total system response. Short-term fluctuations in the array response can be monitored using the internal calibration subsystem, since the transmission properties of the telescope are relatively constant. To date, no such fluctuations have been seen.

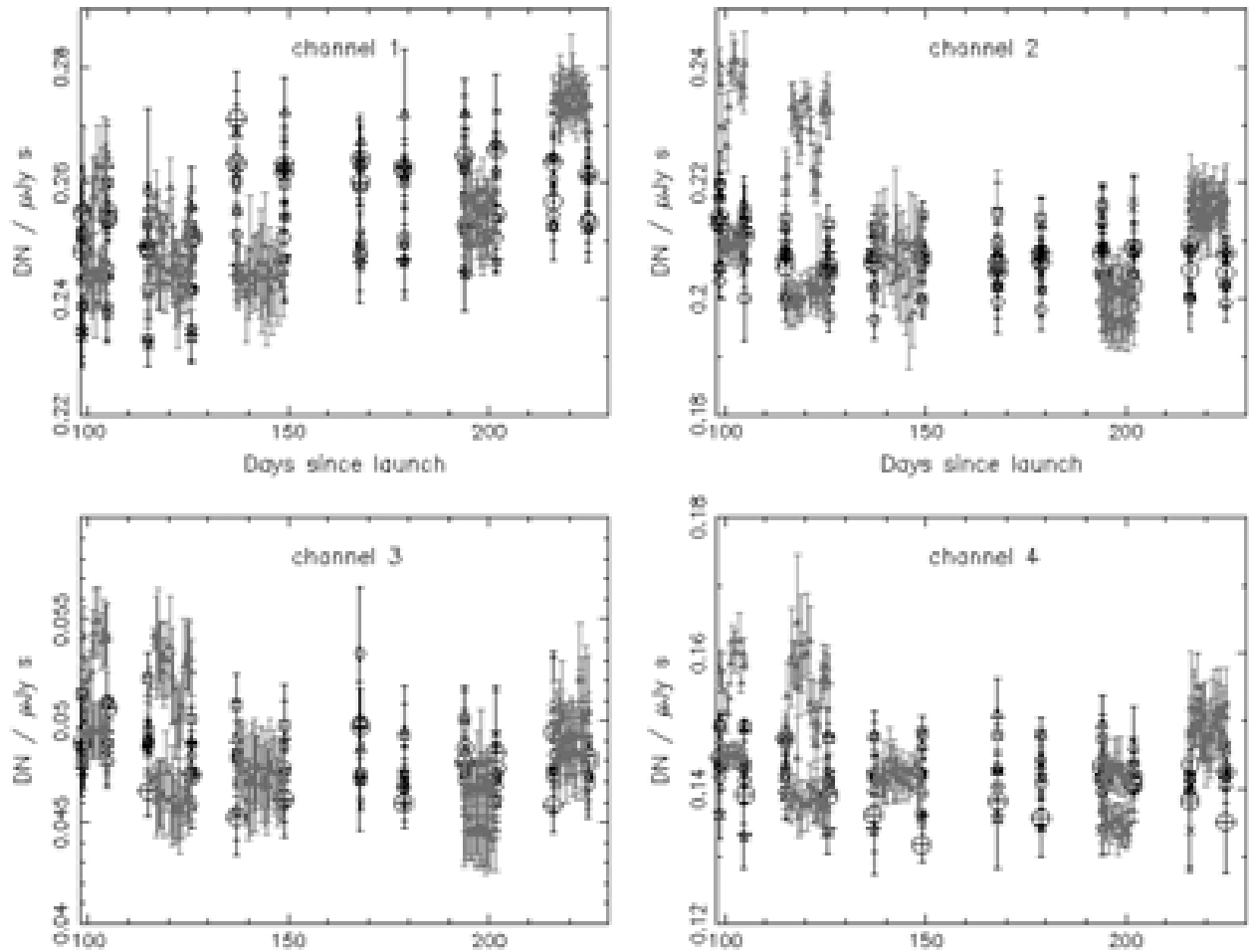

**Figure 10**. The calibration factor DN/µJy/s plotted as a function of days since launch. The primary standards are black, the secondary standards are in gray. Different symbols represent the different stars observed. No significant trend is seen in the ~100 measurements for each star. The measured calibration factors (rms) are 0.255 (0.009) , 0.208 (0.005), 0.048 (0.001 ), and 0.142 (0.004) for channels 1-4, respectively.

Sky flats are obtained using a network of 24 high zodiacal background regions of the sky in the ecliptic plane to ensure a relatively uniform illumination with a reasonable amount of flux. The high-background frames are differenced with frames taken at a low zodiacal background (the field used for measuring the "dark" frames). The data are combined using median filtering to remove the non-uniform component (i.e. stars) and normalized. The resulting flat field (Figure 11) is then divided into the science data in the pipeline.

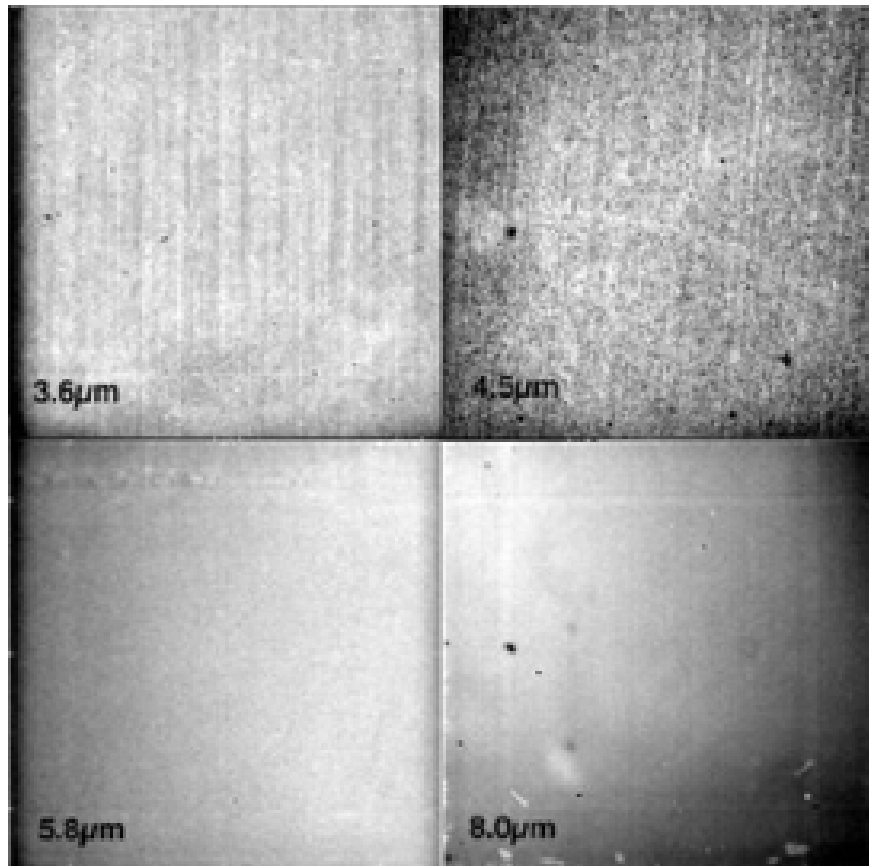

**Figure 11.** IRAC flat field images for the 3.6, 4.5, 5.8, and 8.0 μm channels (channels 1, 2, 3, 4, respectively). Darker represents lower response. The vignetting in channels 1 and 3 is visible in the flats as the dark band at the left edge of those channels. The large black dot in the center-left part of channels 2 and 4 is caused by a dust spot on the pickoff mirror. The "tic-tac-toe" pattern from the scattering of background light from regions just outside the edge of the array is visible in channels 3 and 4 (see Section 3.4).

**5.3. Dark Frames, Offsets and "First Frame" Effect**

The detector dark currents are generally insignificant compared to the sky background. However, there is a significant offset or bias (which can be positive or negative) in a dark frame, and its value must be subtracted from the observations. Especially for shorter frames, the "dark" images are mostly due to electronic bias differences rather than true dark current. Therefore, the number of electrons in a dark image does not scale linearly with exposure time. The shutter has never been used in flight, and therefore isolated dark/bias data cannot be taken. Instead, sky darks are used in conjunction with "lab darks" which were obtained prior to launch when the shutter was used.

As part of routine operations a dark region of the sky near the north ecliptic pole is observed at least twice per campaign, at the beginning and end. These data are reduced and combined in such a way as to reject stars and other astronomical objects with size scales smaller than the IRAC array. An estimate of the zodiacal background is also subtracted from the sky dark. The resulting image of the minimal uniform sky background contains both the bias and the dark current. When subtracted from the routine science data, this eliminates both of these instrumental signatures.

Since the IRAC shutter is open during flight, light constantly falls on the array, even during slews and telemetry downlinks. When IRAC is not taking an exposure, the FPA electronics reset the array every 0.2 sec to avoid having an excess charge build up on the array. If frames are being taken back-to-back, there is only a single reset between frames and the dark frame has a slightly different appearance for the first frame compared to the subsequent ones. In addition, there is a shift in offset level of each frame. Therefore, the dark frames in all four IRAC channels are not constant. A dark image taken with a particular frame time and Fowler number depends on the amount of time elapsed since the previous image and the frame time and Fowler number of the previous image (see Fazio et al.[1] for a discussion of the pixel sampling and Fowler number). In general, the first frame of any sequence of images tends to have a different offset from the others, hence the name "first frame effect." The smallest dark offsets occur when a frame is taken with a very short interval from the preceding image, which occurs when multiple frames are commanded at once. In practice, the offsets vary from one IRAC frame to the next, since many observations take only one image at each sky position, and the time between frames depends on the time to slew and settle from one position to the next. The effect was calibrated on the ground for each pixel, and a correction is applied in the pipeline processing. The effect is largest in channel 3, and small compared to the background in Channel 4. The effect is almost negligible in channel 2, but larger and significant in channel 1 due to an open connection to the clamp drain line on the multiplexer.

### 5.4. Cosmic Ray Effects

The detailed effects of cosmic rays on IRAC images are discussed in detail in another paper[23]. In flight, cosmic rays strike the IRAC detectors at the rate of about one hit per array per second, with each hit affecting about 4 pixels. This was approximately the predicted rate. The most common cosmic rays do not affect the pixel performance in subsequent frames, and are confined to a few pixels around the peak pixel of the hit. The cosmic rays in channel 1 and 2 are more compact than those in channels 3 and 4, which have thicker detectors. In these latter channels, the cosmic ray effects appear as streaks and blobs. However, other events can cause streaks or other multiple pixel structures in the array, and less common energetic and high-Z events can cause muxbleed and residual images in subsequent frames. Most of the cosmic ray effects are removed in the mosaicing process, provided that there are at least three separate observations of each sky position.

## 6. ARTIFACTS DUE TO BRIGHT SOURCES

Array detector artifacts known before launch and discovered in flight include: persistent images in channels 1 and 4, multiplexer bleed, column pulldown, and banding. The known effects were characterized before launch[13,24], and all are described in more detail in the SSC's IRAC Data Handbook[‡]. The impact of these artifacts on the mosaiced image (and the science analysis) can be greatly reduced for observations that are well dithered.

### 6.1. Long-term Persistent Images

Two new long-term persistent image effects[25] were observed in channel 1 and 4 during IOC that were previously not identified during ground tests because of the differences in the way the camera was operated and the illumination field used. Fortunately, both the channel 1 and 4 long-term persistent images can be removed by performing a short thermal anneal where the temperature of the FPA is raised above 23K (Channel 1) or 30K (Channel 4) for two minutes before returning to the nominal operating temperatures of 15K and 6K for channels 1 and 4, respectively.

#### 6.1.1. Channel 1

To perform a data downlink using the high-gain antenna (HGA), the spacecraft must point the HGA towards the Earth, which points the telescope in the opposite direction somewhere in the ecliptic plane. Even though IRAC was not being used to take data during the downlink, it was observed in flight that when a HGA downlink was performed, the dark frames for channel 1 taken even hours later show a residual image of the star field that the instrument was staring at during the downlink. Stars as faint as *K* magnitude of 14 leave residual images above the noise level of the dark frames. The strength of the residual signal in each IRAC pixel is proportional to the logarithm of the incident flux in the field images during the downlink and decays very slowly with time. The non-linearity in the residual image strength causes the residual images to be broadened (FWHM as large as 5-8 pixels) with respect to the IRAC PSF.

---

[‡] http://ssc.spitzer.caltech.edu/irac/

### 6.1.2. Channel 4

Residual images in channel 4 were seen after observing bright objects (*K* magnitude of 4 or brighter). These residual features are extremely long-lived, surviving several weeks and power cycles. The S/N of the residuals tends to be larger at the shorter frame times. The peak S/N of the residuals tends to be below the 3σ noise level of the images even for the brightest stars. The residual images are generated even with short exposures, as in the 12 sec HDR mode. The channel 4 residuals have two separate components: the first is a "negative" component (below the array background level) which decays on timescales of ~1000 sec. The second component is a broadened "positive" residual which survives IRAC power cycles and decays very slowly during each campaign. The positive residuals are much broader and flatter than the unsaturated point sources, but do not have a strong central hole like the one in the saturated stars that generated them.

### 6.2. Multiplexer Bleed, Column Pulldown, Banding

There are several artifacts created in the images when sources on the array approach the saturation level ("bright source"). In the InSb detectors (channels 1 and 2), a multiplexer bleed effect creates a horizontal band that extends across the entire row for several rows above the pixels that contain the bright source (Figure 12). The magnitude of the effect is related to the flux level in each highly illuminated pixel, which for a point source is non-uniform near the peak. Each array has four outputs, and four successive columns are read out simultaneously. This leads to a pattern that repeats every 4 pixels along the row. The muxbleed decays rapidly at first and then almost exponentially as a function of distance from the bright source in the readout direction. As its name implies, the effect is a feature of the multiplexer, and does not represent charge "lost" from the source. Channels 1 and 2 exhibit "column pulldown" (Figure 12), where columns containing a bright source have a lower offset than adjacent columns. Channel 3 exhibits "column pullup", and channel 4 has a "row pullup", where the columns (rows) with a bright source have a higher offset than adjacent ones. These effects are due to the multiplexer, due to FET sources being connected together at various points. Channels 3 and 4 have an additional effect called banding[25] (Figure 13). In these arrays, light is scattered inside the hybridized device, and it reappears all over the array. The scattered light diminishes with distance away from the source, but is enhanced along the row and column containing the source. Post-pipeline processing routines have been developed to remove these artifacts from the IRAC images (Figure 12 and Figure 13).

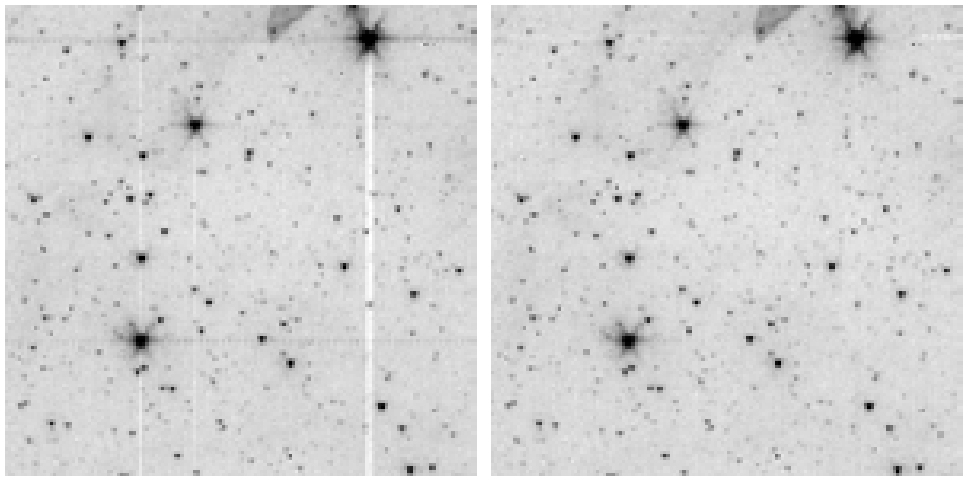

**Figure 12.** The image at left shows a channel 1 pipeline-processed frame, with three stars showing various levels of column pulldown and residual uncorrected muxbleed. The figure at right has had a post-BCD pipeline pulldown and muxbleed correction applied**.** The images are shown as negatives (black is brighter) to enhance the visibility of the artifacts.

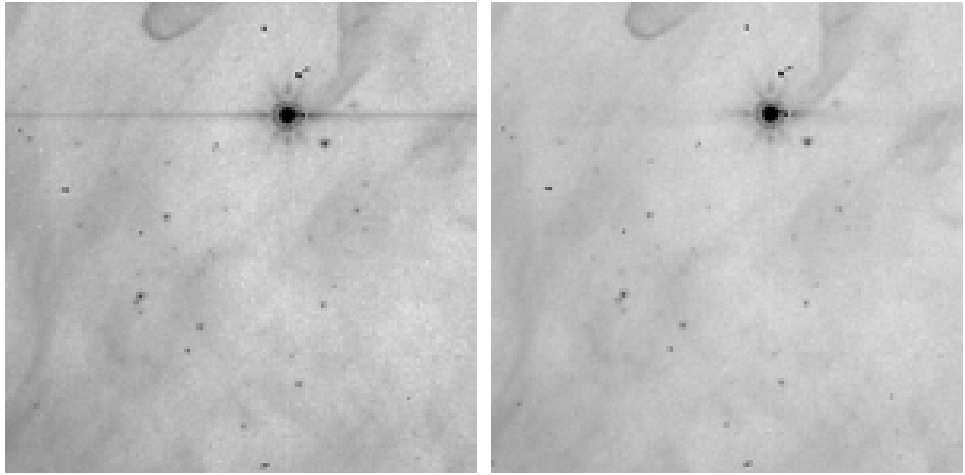

**Figure 13.** The image at left shows a channel 4 pipeline-processed frame containing a bright star with banding. The figure at right shows the frame after a post-BCD pipeline banding correction has been applied**.** The images are shown as negatives (black is brighter) to enhance the visibility of the artifacts.

## 7. SUMMARY

In flight, IRAC meets all of the science requirements established before launch, and in many cases the performance is better than the requirement. The instrument was characterized and focused during the IOC period in the first 90 days after launch, and a baseline calibration was established. Several new array anomalies were observed, and operational techniques were developed to minimize their effect.

IRAC is a powerful survey instrument because of its high sensitivity, large field of view, and four-color imaging in wavelengths of 3.2 to 9.5 μm. IRAC continues to function extremely well since the first in-flight images were produced, seven days after launch, on 2003 September 1.

## 8. ACKNOWLEDGEMENTS

This work is based on observations made with the *Spitzer Space Telescope*, which is operated by the Jet Propulsion Laboratory, California Institute of Technology under NASA contract 1407. Support for this work was provided by NASA through Contract 125790 issued by JPL/Caltech. Support for the IRAC instrument was provided by NASA through Contract 960541 issued by JPL.